\documentclass[3p]{elsarticle}

\usepackage{graphicx, amsmath, amssymb, amsfonts, mathtools, mathrsfs, color}
\usepackage{comment, enumerate, tabularx, multirow}
\usepackage{natbib, hyperref, url}

\newcommand{\abs}[1]{\lvert #1 \rvert}
\newcommand{\omsig}{\sigma_{\omega}}
\newcommand{\omavg}{\omega_0}
\newcommand{\dth}{\Delta \theta}
\newcommand{\dom}{\Delta \omega}
\newcommand{\etas}{\eta_0}

\begin{document}
\title{Anomalous wave statistics induced by abrupt depth change}

\author[Mich]{C.~Tyler Bolles}
\author[GFDI,EOAS]{Kevin Speer}
\author[GFDI,Math]{M.~N.~J.~Moore}

\address[Mich]{Department of Mathematics, University of Michigan, Ann Arbor, MI}
\address[GFDI]{Geophysical Fluid Dynamics Institute, Florida State University, Tallahassee, FL}
\address[EOAS]{Earth, Ocean and Atmospheric Sciences, Florida State University, Tallahassee, FL}
\address[Math]{Department of Mathematics, Florida State University, Tallahassee, FL}


\begin{abstract} 
Laboratory experiments reveal that variations in bottom topography can qualitatively alter the distribution of randomized surface waves. A normally-distributed, unidirectional wave field becomes highly skewed and non-Gaussian upon encountering an abrupt depth transition. A short distance downstream of the transition, wave statistics conform closely to a gamma distribution, affording simple estimates for skewness, kurtosis, and other statistical properties. Importantly, the exponential decay of the gamma distribution is much slower than Gaussian, signifying that extreme events occur more frequently. Under the conditions considered here, the probability of a rogue wave can increase by a factor of 50 or more. We also report on the surface-slope statistics and the spectral content of the waves produced in the experiments.
\end{abstract}
\maketitle

\section{Introduction}

Though once regarded as mythical, rogue waves have now been recorded in oceans across the globe and are no longer doubted as a real phenomenon \citep{kharif2003physical, muller2005rogue, dysthe2008oceanic, ying2011linear, adcock2014physics, hadjihosseini2014stochastic}. Fundamentally, the existence of these large waves is tied to non-normal statistics; if governed by Gaussian statistics, their occurrence would be exceedingly rare and the danger posed modest. From this perspective of anomalous behavior, rogue waves can be considered under the more general framework of turbulent dynamical systems \citep{majda2016introduction, sapsis2013blending, sapsis2013a, sapsis2013b, chen2016filtering, macedo2017universality}. Several physical mechanisms have been shown to produce rogue waves, most notably the Benjamin-Feir (BF) instability that occurs in deep water\footnotemark[1] \citep{benjamin1967disintegration, onorato2001freak, chabchoub2011rogue, viotti2013emergence, adcock2014physics, cousins2015unsteady, chabchoub2016tracking, farazmand2017reduced}, as well as wind excitation \citep{birkholz2016ocean, toffoli2017wind},  opposing currents \citep{onorato2011triggering, toffoli2015rogue}, and geometric ray focusing from 2D bathymetry \citep{heller2008refraction, white1998chance}.

\footnotetext[1]{A mechanism analogous to the BF instability can also arise in optical systems, giving rise to `optical rogue waves' \citep{solli2007optical}.}

	A few recent studies have suggested that anomalous behavior can arise in the much simpler setting of a {\em unidirectional} wave-field propagating over a {\em one-dimensional} variable bottom \citep{sergeeva2011nonlinear, trulsen2012laboratory, gramstad2013freak, viotti2014}. Since these studies are performed outside of the deep-water regime, the BF instability is absent, as are the other mechanisms listed above (i.e.~no wind or current and the bathymetry is strictly 1D). Intriguingly, many of these studies identify certain locations at which the deviation from Gaussianity is maximized and, thus, rogue waves are most likely. Such locations are analogous to the `hot spots' observed in microwave systems \citep{hohmann2010freak}, but without the benefit of 2D geometric focusing.
	
	Inspired by this line of thought, we perform laboratory experiments to examine the statistics of unidirectional waves propagating over a 1D, variable bottom, in the shallow-to-moderate depth regime (outside the influence of the BF instability).  Unlike previous experiments which featured gradual slopes of 1:20 \citep{trulsen2012laboratory}, we focus on abrupt depth transitions. In particular, we consider waves propagating over a step in bottom topography---much like the step potentials considered in quantum mechanics that helped lay the foundation of scattering theory. In accordance with previous results, we find the deviation from Gaussian behavior to be maximized at certain locations. Whereas previous studies only quantified non-Gaussianity in terms of a few statistical moments, we find the complete surface-displacement statistics at these locations to conform closely to a gamma distribution. This clean characterization offers a precise test for theories and may help differentiate the various rogue-wave producing mechanisms.

\section{Experimental setup}
 
\begin{figure}
\begin{center}
\includegraphics[width = 0.85 \linewidth]{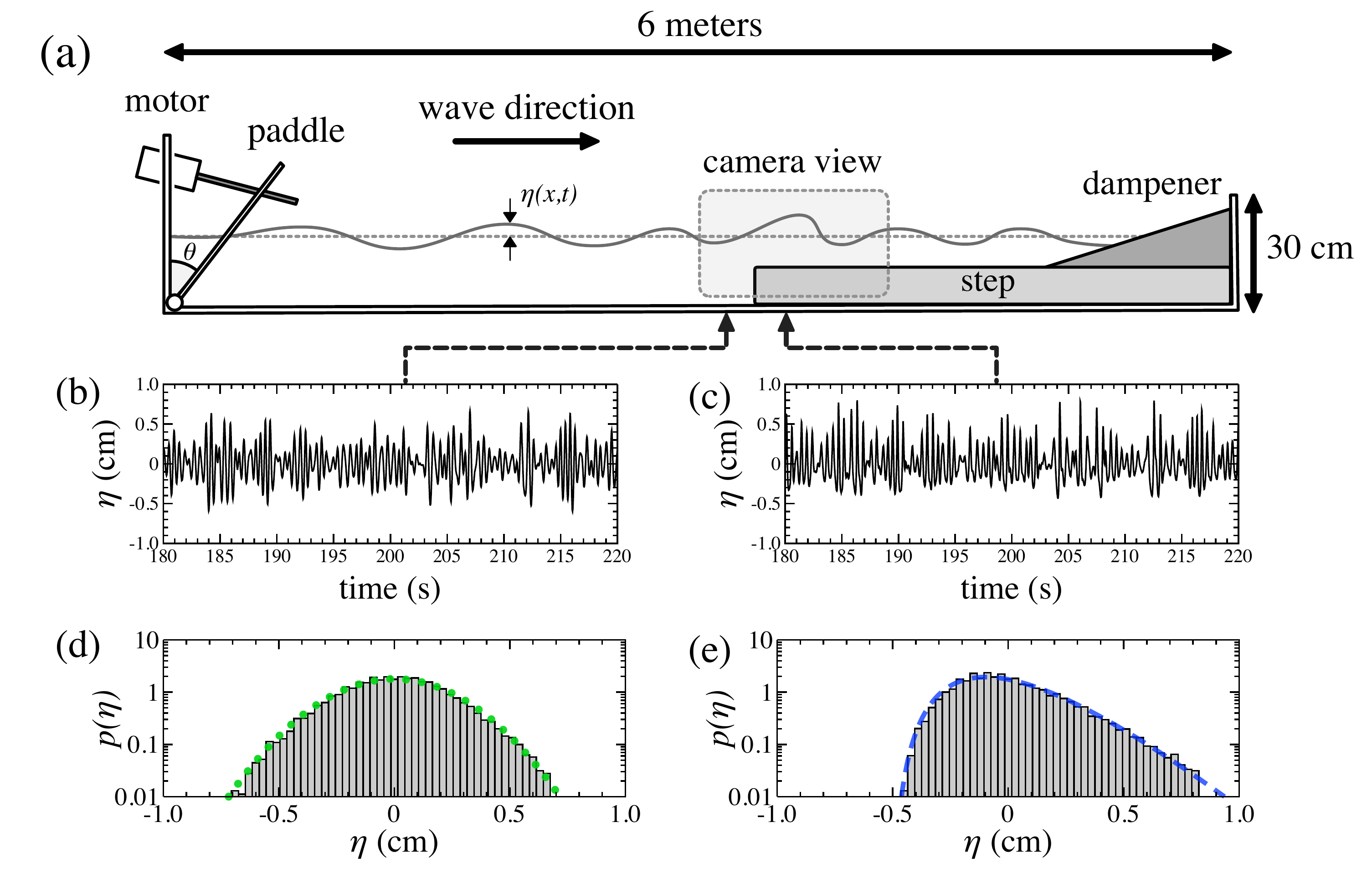}
\caption{(a) Experimental diagram: a pivoting paddle generates randomized surface waves which travel through a 6-meter long wave tank. The waves encounter a depth transition midway through and are damped at the far end.
The free surface is illuminated with LED lights and imaged near the transition.
(b)-(c) Surface elevation measurements taken at a location upstream and downstream of the transition.
(d)-(e) Corresponding histograms. Upstream measurements follow Gaussian statistics (green dotted), whereas  downstream measurements follow a gamma distribution (blue dashed).
}
\label{fig1}
\end{center}
\end{figure}

	Our experiments, diagramed in Fig.~\ref{fig1}(a), consists of a long, narrow wave tank (6 meters in length, 20 cm wide, 30 cm heigh) constructed of transparent plexiglass and filled with water to a depth of 12.5 cm. Waves are generated by a plexiglass paddle (40 cm in length) driven by a 5-phase stepper motor (Oriental CVK569FBK with $0.72^{\circ}$ precision). The paddle is hinged to the bottom so that executes a pivoting motion, which we have found to be highly effective for generating surface waves. A horse-hair dampener located at the far end minimizes reflections back into the tank. To create a depth transition, a plexiglass step (2 meters in length) is inserted at the far end. Spacers allow us to vary the height of the step and thus the ratio of the two depths. In most of the results reported here, the upstream depth is 12.5 cm, the step height is 9.5 cm, and the far-end depth is 3 cm (i.e.~a depth ratio of 0.24)

	To measure the waves, we employ two complementary techniques. Our primary measurements are optical, with video-images taken by a Nikon D3300 from the sideview at 60 fps. The free surface is illuminated by LED lights that run along the bottom of the tank, thus providing the contrast necessary to extract the free surface from images with sub-millimeter-scale resolution (3 pixels per mm) \citep{camassa2012stratified}. The camera is focused on a 48x27 cm window surrounding the depth transition, which, through some trial and error, we determined to be the region of greatest statistical interest. As our secondary measurements, we deploy two AWP-24 depth gauges, enabling somewhat higher temporal resolution (5 ms response time) at fixed locations. We use these gauges primarily for corroboration, as the optical measurements provide the spatial information crucial for identifying regions of statistical anomalies.

 	Central to our study, we aim to create a {\em randomized} incoming wave field so that we can examine how its statistical features are modified by the depth change. We thus prescribe the paddle motion with a pseudo-random signal. Specifically, the paddle angle with the vertical $\theta(t)$ is specified as
\begin{align}
\label{thetaEq}
& \theta(t) = \theta_0 + \dth \sum_{n=1}^N a_n \cos(\omega_n t+\delta_n) \, , \\
\label{anEq}
& a_n = \sqrt{\frac{2 \dom}{\pi^{1/2} \omsig}} \, 
\exp \left( -\frac{(\omega_n - \omavg)^2}{2 \omsig^2} \right) \, , \qquad
\omega_n = \frac{\left( \omavg + 4 \omsig \right) n}{N} \, , 
\end{align}
The mean angle of the paddle is $\theta_0 = 41^{\circ}$ and its motion is given by a superposition of $N$ Fourier modes with (deterministic) Gaussian weights. Importantly, the phases $\delta_n$ are random variables taken from a uniform distribution. In the experiments, we set $\omavg = \omsig = 12.5$ rad/s. Thus, waves are driven with a {\em mean frequency} of 2 Hz and a {\em bandwidth} of 2 Hz. 
The wavelengths $\lambda$ resulting from these forcing parameters can be estimated by the dispersion relation from linear wave theory, $\omega^2 = g k \tanh{kh}$, where $k = 2\pi \lambda^{-1}$ is the wavenumber and $h$ is the depth. The predicted wavelengths are respectively $\lambda = $ 38 cm and 25 cm on the deep and shallow ends ($h = $ 12.5 cm and 3 cm).
We set $N = 3000$, giving a frequency discretization of $3 \times 10^{-3}$ Hz and fundamental period of $T = 300$ seconds. 

	In Eq.~\eqref{thetaEq}, $\dth$ sets the driving amplitude and, more specifically, corresponds to the standard deviation of the paddle angle $\dth^2 ={ \frac{1}{T} \int_0^T \abs{\theta(t) - \theta_0}^2 dt }$.
In the following, we vary $\dth$ in the range $0.125^{\circ}$--$2.0^{\circ}$ to assess how driving amplitude influences wave statistics. This range was selected to probe the various regimes of wave behavior---the lower end produces linear waves, the middle produces weak to moderate nonlinear effects, and the higher end generates strongly nonlinear waves that occasionally break. For $\dth > 2.0^{\circ}$, the dynamics are dominated by breaking and dissipation, and are therefore not discussed here.
In each experimental run, waves are generated for roughly one fundamental period (300 seconds) to allow the system to settle to a quasi-equilibrium before measurements are taken.


\section{Results}

	In the absence of depth variations, the forcing given by Eq.~\eqref{thetaEq} produces {\em normally-distributed} surface waves. This behavior is consistent with linear wave theory and has been verified by control experiments. With a depth transition, however, wave statistics can vary in space and deviate strongly from Gaussian. Figures~\ref{fig1}(b)--(c) show measurements of the surface displacement $\eta(t)$ taken at locations (b) 10 cm upstream and (c) 15 cm downstream of the transition. In this experiment, the driving amplitude is $\dth = 1.38^{\circ}$ and the step height is 9.5 cm, creating a depth ratio of 0.24.
While both  signals in Figs.~\ref{fig1}(b)--(c) exhibit a random character, the upstream measurements fluctuate symmetrical about the mean, whereas the downstream measurements show several spikes biased towards large, {\em positive} displacement. These observation are made more apparent by the corresponding histograms in Figs.~\ref{fig1}(d)--(e). Indeed, the upstream measurements are distributed symmetrically about the mean and, in fact, follow a Gaussian distribution closely (green dotted curve), as is consistent with linear wave theory. The downstream measurements, however, skew heavily towards the right, or positive displacement, with the mean remaining $\eta =0$. The deviation from normal statistics signals nonlinear effects. We have found this histogram to conform closely to a gamma distribution (blue dashed curve), defined as
\begin{equation}
\label{gamma}
p(\eta) = \frac{e^{-\alpha}}{\etas \Gamma(\alpha)} 
\left( \alpha + \eta/\etas \right)^{\alpha-1} \exp{ \left(- \eta/\etas \right)} \, , \quad \text{for } 
-\alpha \etas \le \eta < \infty  \, .
\end{equation}	
The above is a gamma distribution with mean zero, scale parameter $\etas$, and shape parameter $\alpha$, where the latter two parameters are fit to obtain the blue curve in Fig.~\ref{fig1}(e). The agreement between the measurements and fit is remarkably close over nearly two orders of magnitude ($ 0.02 \le p(\eta) \le 2$). Compared to a Guassian, the relatively slow decay rate of the gamma distribution indicates an increased frequency of high wave crests (large $\eta$).

\begin{figure}
\begin{center}
\includegraphics[width = 0.80 \linewidth]{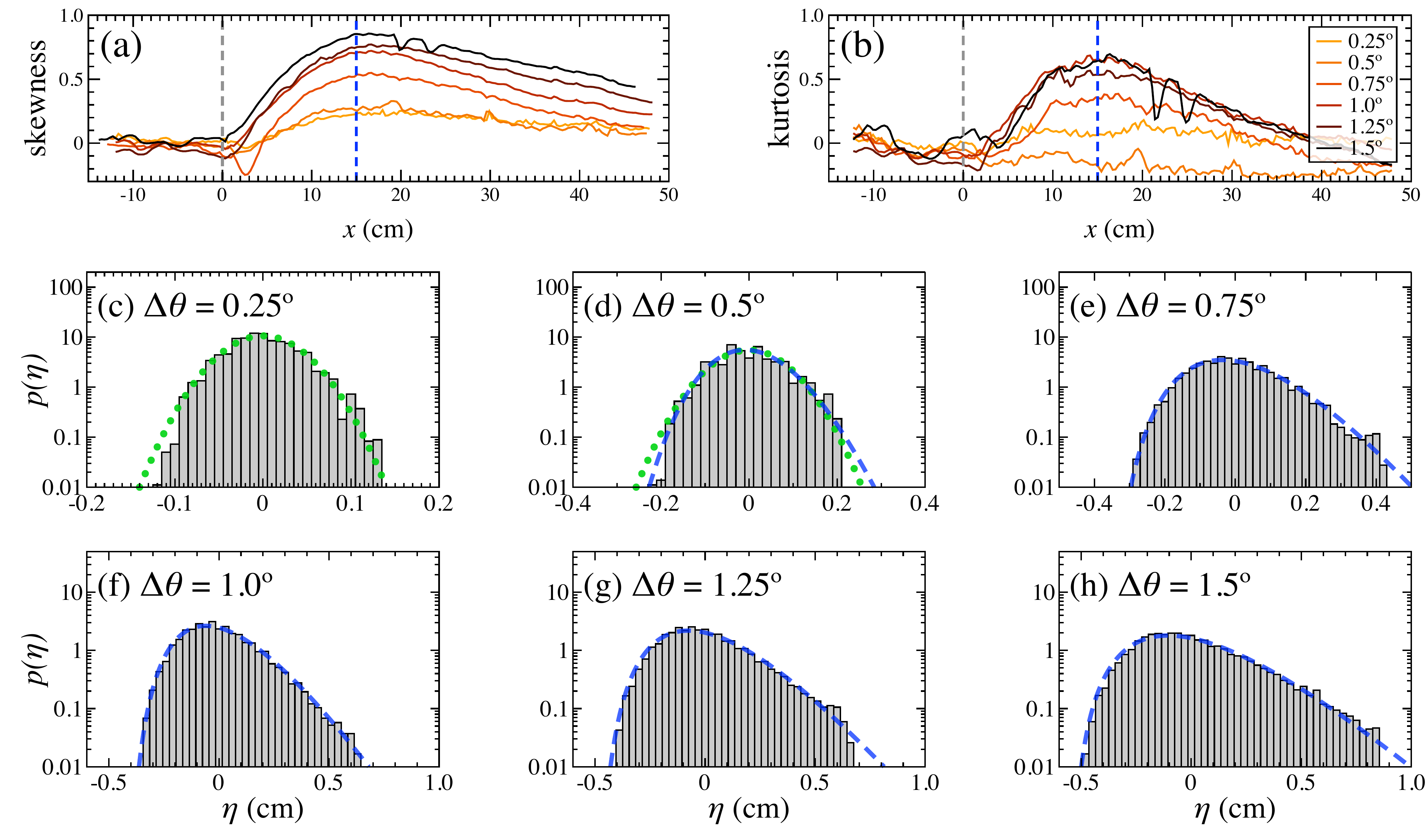}
\caption{Ensemble of experiments with different driving amplitudes:
(a)--(b) Skewness and kurtosis plotted against distance $x$ from the depth transition for 6 different driving amplitudes (see legend). Both skewness and kurtosis reach a maximum 15 cm downstream of the transition (blue vertical line).
(c)--(h) Histograms of $\eta$ taken at $x = 15$ cm for the same 6 driving amplitudes. For larger amplitudes, $\dth \ge 0.75^{\circ}$, the histograms follow a gamma distribution (blue dashed curves).
}
\label{fig2}
\end{center}
\end{figure}
 
	The fixed-location measurements indicate very different behavior on either side of the depth transition, but provide little detail on how wave properties vary in space. Fortunately, the optical measurements allow us to extract statistical information {\em continuously} in space, rather than at a discrete set of positions. This capability is crucial for identifying regions of anomalous activity, as previous studies have suggested such regions to be highly localized \citep{viotti2014}. While examining the complete distribution of $\eta$ at a continuum of positions would prove impractical, the first few statistical moments provide useful proxies. In particular, we focus on the {\em skewness} and {\em excess kurtosis} (hereafter simply called kurtosis), which are normalized and shifted third and forth moments respectively. Since both quantities are zero for a normal distribution, they indicate how wave statistics deviate from Gaussian.
	
	In Figs.~\ref{fig2}(a)--(b), we show the skewness and kurtosis as they vary in space for an ensemble of experiments. Here, $x$ is the distance from the depth transition, where negative $x$ is upstream and positive $x$ downstream. The figure shows 6 different driving amplitudes in the range $\dth = 0.25^{\circ}$--$1.5^{\circ}$ (colored yellow to black). In all cases, both moments are essentially zero for $x<0$, suggesting that wave statistics are nearly Gaussian upstream of the transition. For $x>0$, however, the experiments differ. In the smaller-amplitude experiments ($\dth \le 0.5^{\circ}$) skewness and kurtosis remain small (albeit with fluctuations), but in the larger-amplitude experiments ($\dth > 0.5^{\circ}$) both moments grow substantially and reach a   maximum somewhere downstream. Skewness and kurtosis then decay farther downstream, presumably due to nonlinear dissipation. Remarkably, the location of the maximum, $x = 15$ cm, is nearly the same for both skewness and kurtosis and for all of the amplitudes tested. This consistency indicates a small region of highly intensified wave activity, whose location is independent of driving amplitude.
Both the location and size of the anomalous region are on the scale of the wavelengths, 25--38 cm, induced by the dominant 2 Hz forcing (as predicted by linear dispersion).
 
	With the anomalous region identified, we now examine how the non-Gaussian features change with driving amplitude. In particular, we aim to test the robustness of the gamma-distribution description. Accordingly, Figs.~\ref{fig2}(c)--(h) show histograms of $\eta$, all taken at the anomalous location $x = 15$ cm, for the six driving amplitudes discussed above. The first two histograms, (c) and (d), deviate only slightly from Gaussian (green dotted curves), with the second showing signs of transitioning towards a skewed distribution. The next four histograms with larger driving amplitude, (e)--(h), all skew heavily towards the right, or positive surface displacement. Indeed, each of these histograms conforms closely to a gamma distribution (blue dashed curves), with different values of $\etas$ and $\alpha$. Notice that, as the driving amplitude increases from $\dth = 0.75^{\circ}$ to $1.5^{\circ}$, the skewness of the distribution becomes more pronounced, indicating that extreme wave crests occur more frequently.
 
\begin{figure}
\begin{center}
\includegraphics[width = 0.95 \linewidth]{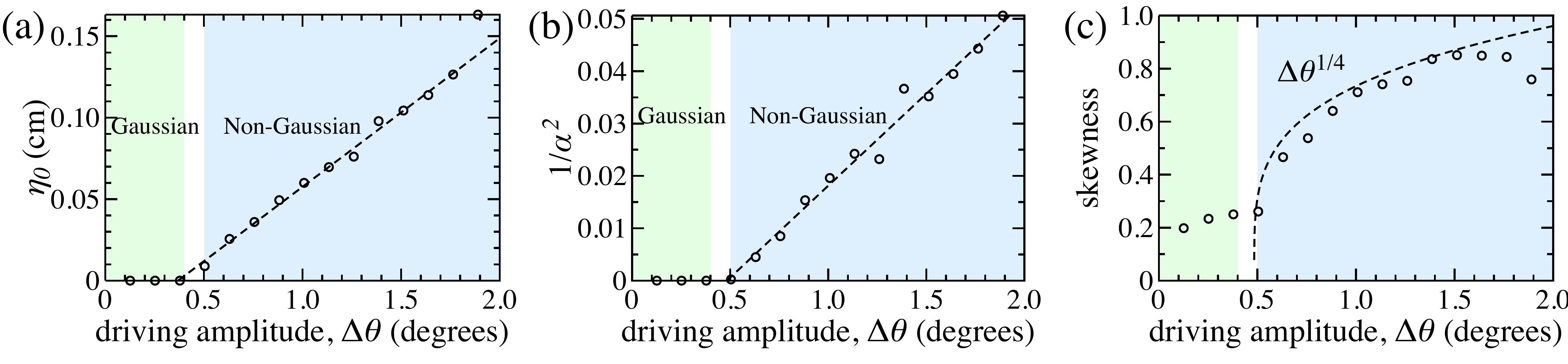}
\caption{(a)--(b) Analysis of the scale $\etas$ and shape $\alpha$ parameters over an ensemble of 15 experiments. Small amplitudes generate Gaussian statistics (green region), while larger amplitudes give rise to the gamma distribution (blue region). 
For $\dth \ge 0.5^{\circ}$, both $\etas$ and $\alpha^{-2}$ grow linearly with $\dth$. 
(c) The linear fit of $\alpha^{-2}$ enables a scaling-law prediction for skewness, which compares well with direct measurements.
}
\label{fig3}
\end{center}
\end{figure}
 
	These figures indicate that, within the anomalous region, the {\em complete} wave statistics can be described by only two parameters, $\etas$ and $\alpha$. What physical meaning do these parameters carry? The first, $\etas$, simply sets a length scale of the gamma distribution, while the second, $\alpha >1$, controls its shape: large $\alpha$ signifies a nearly symmetric distribution and smaller $\alpha$ a highly skewed one. By analyzing how these parameters change with experimental conditions, we can understand the corresponding changes in anomalous wave activity. Accordingly, Figs.~\ref{fig3}(a)--(b) show the values of these parameters extracted from 15 different experiments, with the driving amplitude systematically increased from $\dth = 0.125^{\circ}$ to $1.875^{\circ}$. For very small amplitude, the statistics are nearly Gaussian (green region). Above a threshold of about $\dth = 0.5^{\circ}$, however, the statistics deviate from Gaussian and instead follow the gamma distribution (blue region). In this regime, $\etas$ increases and $\alpha$ decreases with driving amplitude, meaning that, not only is the distribution growing in length scale but it is also becoming increasingly skewed. More precisely, we have found the growth of $\etas$ and $\alpha^{-2}$ to be nearly linear with $\dth$ (dashed lines in Figs.~\ref{fig3}(a)--(b)). The linear growth of $\etas$ might have been anticipated, as it is plausible that the driving amplitude should directly set a length scale for the surface-displacement distribution. We have no such explanation for the linear growth of $\alpha^{-2}$, however, and simply report it as an experimental finding.

	The clean characterization of the {\em entire} surface-displacement distribution via only two parameters allows one to immediately predict any statistical feature of $\eta$, for example its moments. In particular, the gamma distribution, Eq.~\eqref{gamma}, has a skewness of $2 \alpha^{-1/2}$. The observed linear growth of $\alpha^{-2}$ with driving amplitude, $\dth$, thus implies the scaling law: skewness $\sim \dth^{1/4}$. In Fig.~\ref{fig3}(c) we show, for each of the 15 experiments, the skewness taken directly from the measurements (circles), along with this scaling prediction (dashed curve). The scaling law accounts for the experimental trend remarkably well within the regime of anomalous activity. We note that in obtaining this prediction, the formula $2 \alpha^{-1/2}$ was applied directly to the linear estimate of $\alpha^{-2}$ from Fig.~\ref{fig3}(b), with no additional fitting parameters introduced. A similar scaling estimate is possible for other moments, for example kurtosis, though the direct calculation of higher moments from experimental data becomes increasingly sensitive to noise, and we therefore limit attention to skewness here.
	
\begin{figure}
\begin{center}
\includegraphics[width = 0.85 \linewidth]{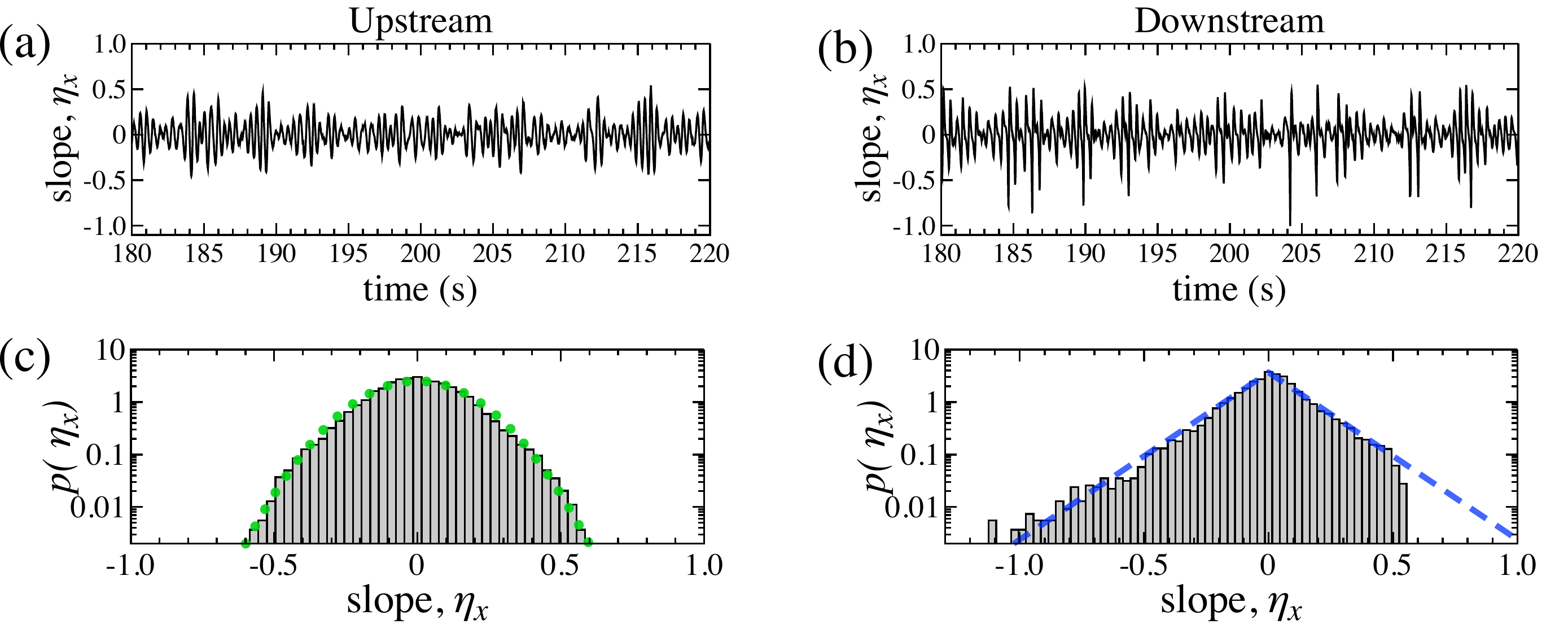}
\caption{Surface-slope statistics. (a)--(b) The time series of the slope, $\eta_x$, taken at a location (a) 10 cm upstream and (b) 15 cm downstream of the depth transition. (c)--(d) Corresponding histograms. The upstream slopes are nearly Gaussian (green dotted), while the downstream slopes are well fit by a Laplace distribution (blue dashed), preferentially truncated on the left.
}
\label{fig4}
\end{center}
\end{figure}
 
	In addition to large displacements, anomalous waves often exhibit extreme {\em surface slopes}. A major advantage of optical measurements is the ability to extract the slope, $\eta_x$, through numerical differentiation. Though subject to higher noise sensitivity (which can be mitigated somewhat by smoothing), these measurements offer insight into the waveforms associated with non-Gaussian statistics. In Figs.~\ref{fig4}(a)--(b), we show time series of the surface slopes, $\eta_x$, extracted from the same experiment shown in Fig.~\ref{fig1}. As before, the data are taken at a location (a) 10 cm upstream and (b) 15 cm downstream of the depth transition. While the upstream slope measurements show relatively small, symmetric fluctuations about the mean, the downstream measurements show larger events biased towards {\em negative} slope, $\eta_x < 0$. Since waves travel in the positive-$x$ direction, these events indicate waveforms with a steep leading surface and more gradual trailing surface. The corresponding histograms are shown in Figs.~\ref{fig4}(c)--(d). Much like displacement, the slope measurements follow a Gaussian distribution (green dotted) upstream of the transition, but deviate markedly from Gaussian downstream. In particular, we have found the downstream data to be partially described by a Laplace distribution (blue dashed curve)
\begin{equation}
p(\eta_x) = \frac{1}{2 s_0} \exp (- \abs{\eta_x} / s_0) \, ,
\end{equation}
where $s_0$ sets a scale of the slope and is a fit parameter. Notice that the Laplace distribution is symmetric about $\eta_x = 0$, even though we observed a clear bias towards negative slopes in the downstream data. How can these two observations be reconciled? Curiously, the histogram in Fig.~\ref{fig4}(d) appears to be truncated preferentially on the right side. Thus, the downstream slope measurements follow a Laplace distribution fairly well, with the exception that events of large, positive slope seem to be omitted. Similar to the gamma-distribution fit, we have found this Laplace-distribution description of the slope to be reasonably robust to changes in driving amplitude and step height, though for brevity, we do not detail those results here.

\begin{figure}
\begin{center}
\includegraphics[width = 0.8 \linewidth]{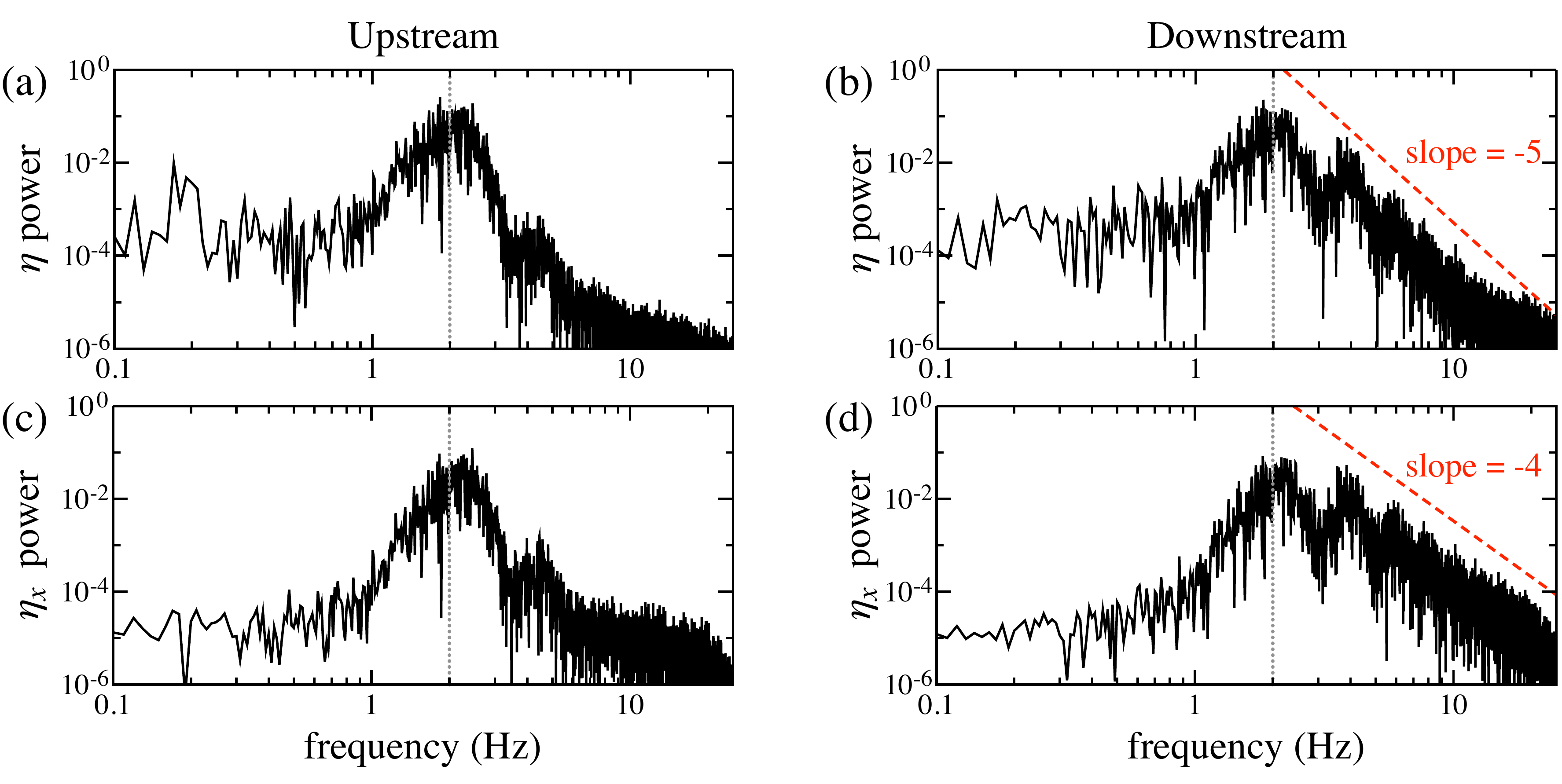}
\caption{Spectral content of surface waves. (a)--(b) Power spectrum of surface displacement, $\eta$, taken at a location 10 cm upstream and 15 cm downstream of the depth transition, respectively, from the same experiment detailed in Figs.~\ref{fig1} and \ref{fig4}.
(c)--(d) The same for surface slope, $\eta_x$.
The upstream spectra decay rapidly with frequency, whereas the downstream spectra decay more gradually. Estimated power laws are shown by the red dashed lines.
}
\label{fig5}
\end{center}
\end{figure}
 
	Lastly, we briefly report on the spectral content of the waves generated in our experiments, in particular how the spectrum is modified by the depth transition. Figure \ref{fig5}(a)--(d) shows the power spectrum of the displacement, $\eta$, and slope, $\eta_x$, taken from the same experiments featured in Figs.~\ref{fig1} and \ref{fig4} and at the same two locations (10 cm upstream and 15 cm downstream of the transition). In the upstream measurements, the power spectra of both $\eta$ and $\eta_x$ peak near the mean-forcing frequency of 2 Hz (faint vertical line), then decay rapidly to a noise-level of about $10^{-5}$. The downstream measurements also peak around 2 Hz, but decay more gradually at high frequencies. In particular, both spectra appear to decay algebraically, with powers estimated as -5 for displacement and -4 for slope (red dashed lines). These relatively slow decay rates indicate that, within the anomalous region, waves possess an elevated level of high frequencies. We comment that the decay rate of $\eta$ is consistent with the Phillips spectrum \citep{phillips1958equilibrium}, which has also been observed in previous experiments \citep{nazarenko2010statistics} and numerical simulations \citep{viotti2014}.

\section{Conclusions}	

These experiments reveal some basic, quantitative information on the emergence of anomalous waves from changes in bottom topography. We have found the deviation from Gaussian behavior to be maximized a short distance downstream of a depth transition. At such locations, wave statistics conform closely to a gamma distribution. Compared to a Gaussian, the relatively slow decay of the gamma distribution indicates an increased likelihood of extreme event, e.g.~rogue waves. The emergence of the gamma distribution appears to be robust to changes in parameters, for example driving amplitude and depth ratio, once a threshold amplitude is exceeded. 
As opposed to characterizing a distribution in terms of a few of its statistical moments, the gamma distribution appears to provide a near-complete description of surface-displacement statistics (over roughly two decades of $p(\eta)$), with only two parameters that need to be estimated.  Additional findings on the surface-slope distribution and power spectra point to waves of steep leading surface and higher-frequency content within the anomalous region. These findings, taken together, offer a stringent test for anomalous-wave theories, and may help guide their further development.	

While we recognize the limitations of these idealized experiments, many of the basic results may apply more generally to naturally occurring waves. For narrow-band forcing, a normal distribution of surface displacement produces a Rayleigh distribution of wave maxima, which has been found to agree generally with ocean observations. But circumstances of enhanced nonlinearity, for example due to variable bathymetry or wind excitation, can produce non-normal statistics with a greater number of extreme waves than would be expected from the Rayleigh distribution. Interestingly, recent measurements taken in the Sea of Japan exhibit exponential tails \citep{hadjihosseini2014stochastic}, which are comparable to those observed in our experiments and suggests similar principles may be at work.

We close with a simple calculation for the probability of a `rogue wave' implied by the distributions observed in our experiments. Though conventions vary somewhat, let us define a rogue wave as one having a crest that exceeds 4 standard deviations of the surface displacement, i.e.~$P({\eta} > 4 \sigma \, | \, \eta >0)$. For a normal distribution, this definition gives a rogue-wave  probability of $6.3 \times 10^{-5}$, equivalent to most other definitions in the literature\footnotemark[2]. Non-normal distributions, however, may yield different probabilities. For the gamma distribution Eq.~(\ref{gamma}), in particular, the probability $P({\eta} > 4 \sigma \, | \, \eta >0)$ depends only on the shape parameter, $\alpha$. The measurements in Fig.~\ref{fig3}(b) indicate a typical value of $\alpha = 6.5$ and, in the most extreme case, $\alpha = 4.5$. For these two values, the probability of a rogue wave increases by a factor of 47 and 65, respectively, over a Gaussian distribution, indicating a far greater prevalence of these extreme events.

\footnotetext[2]{Most commonly, rogue waves are defined as the maximum crest-to-trough height exceeding some multiple (often 2 or 2.2) of the significant wave height $H_s = 4 \sigma$, where $\sigma$ is the standard deviation of $\eta$. However, we prefer a definition that depends on the surface displacement only, since that quantity can be measured directly in our experiments without the ambiguity and potential statistical bias associated with identifying crests/troughs of randomized wave trains.
}

\section*{Acknowledgements}
We would like to acknowledge Daniel Kuncicky, Anthony Diaz, and Robert Broedel for valuable assistance in designing, constructing, and operating the experimental apparatus.
We would like to thank Andrew J. Majda and Nan Chen for enlightening and inspirational discussions. 
CTB acknowledges support from the IDEA grant at Florida State University, as well as from the Geophysical Fluid Dynamics Institute. MNJM acknowledges support from the Simons Foundation (award ID 524259). KS acknowledges support from NSF OCE 1231803 and NSF OCE-1536045.

\bibliographystyle{jfm}
\bibliography{wavesbib}

\end{document}